\documentclass[7pt]{article}

\usepackage{longtable}
\begin{document}

\title{Pressure effects on the Raman spectrum of $CaZnF_4$}

\author{Qiuping A. Wang \\ Institut Sup\'erieur des Mat\'eriaux du Mans, \\
44, Avenue F.A. Bartholdi, 72000 Le Mans, France \and Sophie Sala\"un \\ Laboratoire de Physique de
l'Universit\'e de Bourgogne, CNRS UMR 5027, \\ Facult\'e des Sciences Mirande, 21078 Dijon Cedex, France
\and Alain Bulou \\ Laboratoire de Physique de l'\'Etat Condens\'e, CNRS UMR6087, \\
Universit\'e du Maine, 72085 Le Mans Cedex 9, France}

\date{}

\maketitle

\begin{abstract}
The pressure influence on the lattice vibration of $CaZnF_4$ has been studied by Raman diffusion up to 17 GPa.
Most Raman frequencies increase with increasing pressure. Three singularities in the pressure induced frequency
evolution are observed around 1.5 GPa, 10 GPa and 17 GPa. The samples pressurized to 17 GPa or higher do not
revert to the ambient pressure phase after being released, the new phase showing different Raman spectra from
the ordinary one. It is suggested that $CaZnF_4$ undergoes probably sudden lattice deformations at about 1.5
GPa and 10 GPa, and an irreversible phase transformation above 17 GPa.
\end{abstract}

Keywords: Pressure, lattice dynamics, Raman scattering, Phase transition.

{\small PACS :  64.70.Kb, 62.50.+p; 78.30.Hv}


\section{Introduction}
$CaZnF_4$ belongs to a family of crystals with tetragonal scheelite structure (Figure 1) which is known to be
one of the crystal families undergoing ferroelastic phase transition with temperature and
pressure\cite{Bulo1,Wada,Pinc1,Pinc2,Davi,Benu}. Recently, the lattice dynamics of an important member of this
family, $YLiF_4$, a laser host, has been studied theoretically and experimentally with IR absorption, Raman and
neutron scattering\cite{Sophie,Mill,Forn1,Sen}, as a function of both temperature (from 10 K to 1000 K) and
pressure (up to 40 GPa)\cite{Saran,Liu,Wang1}. These works, especially those under high pressures, are useful
for understanding the dynamics of this lattice and for the verification of the validity of the rigid ion models
employed in the dynamical calculations\cite{Sophie,Sen}.

Although no phase transition was observed with temperature, $YLiF_4$ has shown many Raman and luminescence
(from $Pr^{3+}$ and $Eu^{3+}$ doping ions) anomalies with increasing pressure which are not easy to interpret
with only the stiffening effect of the lattice. So it has been suggested that the compounds undergoes several
lattice transformations or distorsions due to high pressure\cite{Liu,Wang1}, especially when the latter is of
the same order as the elastic constants $C_{11}$ and $C_{33}$ ($\sim 100$ GPa) of the
lattice\cite{Sophie,Blan}. Nevertheless, for want of further experimental evidences, the above conjectures have
been made with precaution.

Another member of the family is $CaZnF_4$ of which the lattice dynamics is well studied\cite{Sophie}. Having
$C_{4h}^6$ space group, this crystal gives very intense Raman scattering, offering an excellent sample for high
pressure Raman scattering. According to dynamical analysis, there are 13 active Raman modes ($3A_g\oplus
5B_g\oplus 5E_g$). Some of the authors of present paper have reported the Raman frequencies from 300 K to 10 K
and shown that the crystal structure, as that of $YLiF_4$, is stable in that temperature range\cite{Sophie}.

In this paper, we present the results of a Raman study of $CaZnF_4$ under high pressure up to 26 GPa. Several
anomalies in Raman spectra and our analysis will be reported.

\subsection{Experiment and result}
The samples are small single crystals of about 100 $\mu$m width and 50 $\mu$m thick. The pressure was generated
with a gasketed diamond anvil cell Diacell MK3 with a pair of anvils of type IIb/a diamond. The anvil flats are
550 $\mu$m in diameter. The cylindrical pressure chamber is 100 $\mu$m in diameter and less than 100 $\mu$m
high. A chip of the sample is put into the pressure chamber together with a small ruby (for pressure
measurement) and the pressure medium consisting of a 4:1 mixture of methanol:ethanol. Pressure calibration is
carried out by the R lines of ruby with the non-linear pressure scale\cite{Mao}. The Raman spectra are excited
by the 514.5 nm line of an argon ion laser with 130 mW total power and collected with a Dilor-Z24
triple-monochromator single channel Raman spectrometer. Different orientations of the sample in the pressure
chamber are taken in order to obtain as many Raman lines as possible, polarization analysis being difficult in
the pressure cell.

\begin{longtable}[b]{llllll}
\caption{Raman frequencies $\omega$ observed under ambient conditions and their temperature and pressure
dependence $d\omega/dT$ (0-300 K), $(d\omega/dP)_1$ (0-1.5 GPa) and $(d\omega/dP)_2$ (1.5-7.0 GPa). The atom
movements of the lattice vibration are given approximately according to the notations
of Miller et al\cite{Mill}.}\label{tab1}\\
\hline mode & $\omega$ & $d\omega/dT$  & $(d\omega/dP)_1$ & $(d\omega/dP)_2$ & movements of \\
            &{\scriptsize ($cm^{-1}$)}& {\scriptsize ($cm^{-1}$)/K} & {\scriptsize ($cm^{-1}$)/GPa} & & lattice vibrations \\
\hline Bg &  82 & 0      & 1.33 & 1.45     & $F_4``\nu_2"$+z translation \\
\hline Eg & 128 & -0.010 & 6.67 & 0.80     & $F_4``\nu_4"$+Ca-xy trans.\\
\hline Ag & 138 & -0.007 & 1.33 & 0.71     & $F_4``R_z"+F_4``\nu_2"$ \\
\hline Eg & 194 & -0.003 & 1.21 & $\sim 0$ & $F_4``\nu_4"$+Ca-xy trans. \\
\hline Ag & 202 & -0.021 & 5.33 & 2.00     & $F_4``\nu_2"+``R_z"$ \\
\hline Bg & 208 & 0      & 1.31 & 4.02     & $F_4``\nu_2"+``\nu_4"$+z-trans.\\
\hline Eg & 248 & -0.014 & 6.67 & 5.09     & $F_4``\nu_4"$+xy-trans. \\
\hline Bg & 276 & -0.007 & 6.67 & 6.18     & $F_4``\nu_2"$+z-trans. \\
\hline Eg & 298 & -0.017 & 2.50 &          & $F_4``\nu_4"$+Ca-xy trans.\\
\hline Bg & 318 & -0.021 & 2.99 & 6.55     & $F_4``\nu_2"+``\nu_4"$+Zn-z trans.\\
\hline Bg & 432 & -0.007 & ?    & $\sim 0$ & $F_4``\nu_4"$+z-trans.+Zn-z trans.\\
\hline Eg & 468 & -0.017 & -1.02& $\sim 0$ & $F_4``R_x(R_y)"$+Zn-xy trans.\\
\hline Ag & 474 & -0.017 & 5.91 & 6.00     & $F_4``\nu_1"$ (stretching) \\
\hline
\end{longtable}

Several Raman spectra at different pressures are shown in Figure 2. All 13 expected lines are observed under
pressure. The Raman frequencies are reported in Table \ref{tab1} which also shows that, in general, the
influence of temperature is not equivalent to that of pressure. Small (or great) slope $d\omega/dT$ sometimes
corresponds to great (or small) $d\omega/dP$.

\subsection{Discussion}
The hydrostatical pressure mainly leads to contraction of the lattice, inversely proportional to the elastic
constants $C_{11}$ and $C_{33}$ which are 121 GPa and 156 GPa, respectively, for $YLiF_4$. So for a pressure of
about 10 GPa, It can be estimated that the lattice deformation is roughly $8\%$ along the parameter \textbf{a}
and $6\%$ along the parameter \textbf{c}. This estimation should also apply for $CaZnF_4$. On the other hand,
it appears from figure 1 that the easiest way for the decrease of \textbf{a} is to increase the angle of the
rotation of the $ZnF_4$ tetrahedra around the tetragonal axis up to $45^o$. So it is expected that increasing
pressure may first of all result in a rotation of $ZnF_4$ around \textbf{c}.

Now if we look at the interatomic distances\cite{Sophie}, we note that the distances $d_{Zn-F}=1.931$ \AA \,
(with force constant $A_{Zn-F}\sim 190 N/m$) and $d_{Ca-F}=2.336$ \AA \, (with force constant $A_{Ca-F}\sim 70
N/m$), while the sum of the ionic radii $r_{Zn}+r_F=1.93$ \AA \, and $r_{Ca}+r_F=2.45$ \AA. On the other hand,
the shortest $d_{F-F}$ is 3.064 \AA \, in the tetrahedra and 2.755 \AA\, between the tetrahedra while the sum
$r_F+r_F=2.38$ \AA\, (with force constant $A\sim 10 N/m$). So it seems that, under pressure, a distorsion of
the tetrahedra $ZnF_4$ and the polyhedra $CaF_8$ would be easier than the contraction of the latters. In this
sense, $CaZnF_4$ is completely different from $YLiF_4$ due to the interatomic distances\cite{Wang1} and the
force constants $A_{Li-F}\sim 60 N/m$ and $A_{Y-F}\sim 130 N/m$\cite{Sophie,Blan}. So the $ZnF_4$ tetrahedron
is much more rigid than the $LiF_4$ one. However, a common behavior of these two lattices is the possible
pressure induced rotation of the tetrahedra around \textbf{c}.

{\bf Pressure dependence of Raman modes} : From Figure 3, we see that the main effect of pressure is the
increase of most of the observed frequencies. The mode Bg at 432 $cm^{-1}$ undergoes frequency decrease but it
was determined with ambiguity due to its weak Raman intensity. The frequency of the mode Eg at 468 $cm^{-1}$
remains almost constant in the whole pressure range. The pressure dependence $(d\omega/dP)$ is given in Table
\ref{tab1}. We note a slope change for some frequencies in the vicinity of 1.5-2 GPa. To illustrate this, the
slope $d\omega/dP$ is given for two pressure ranges : 0-1.5 GPa and 1.5-7.0 GPa. For some modes, the pressure
slopes can be extended to higher pressures. The corresponding atomic movement of each mode is described in
Table 1 with the notations of symmetry coordinates given in \cite{Mill}.

{\bf Singularities around 1.5 GPa} : This singularity mainly consists of the change in $d\omega/dP$ and of the
disparition of two lines at about 140 $cm^{-1}$ (Ag) and 310 $cm^{-1}$ (Eg). An interesting point here is that
the two highest frequencies, Ag (474 $cm^{-1}$) and Eg (468 $cm^{-1}$), separate from each other very rapidly,
Ag increases but Eg remains constant up to very high pressure. From Table 1, we see that this Ag mode is a
stretching ($F_4``\nu_1"$) of $ZnF_4$ tetrahedron, while this Eg mode is rather a distorsion of $LiF_4$
tetrahedron along \textbf{c} and the translation of $Zn$ on the (001) plane. If we think of the rigidity of
$ZnF_4$ tetrahedron, the strong increase of the stretching frequency is logical in the whole pressure range.
The relative independence of this Eg mode from pressure can also be understood if we note that the pressure
induced contraction of the lattice is relatively small along \textbf{c}. As a matter of fact, other Eg modes
also have small slope above 1.5 GPa, apart from that at 248 $cm^{-1}$ which principally corresponds to a
deformation of $ZnF_4$ tetrahedron on the (001) plane\cite{Sophie}, so that its frequency strongly increases
due to the important contraction of \textbf{a} under pressure.

Similar phenomenon has been observed for $YLiF_4$ in our previous work\cite{Wang1} around 6 GPa. This Raman
behavior has been interpreted as a result of the lattice stiffening due to increasing pressure. We think that
this is not sufficient to account for the rapid slope change at this pressure. A plausible conjecture is that
the tetrahedra, in order to relax constraints, begin to rotate around \textbf{c} at certain pressures without
significant structure or symmetry change. This rotation up to $45^o$ (see figure 1c) takes place at a higher
pressure in $YLiF_4$ due to the flexibility of its $LiF_4$ tetrahedra than in $CaZnF_4$ which has very rigid
$ZnF_4$ tetrahedra and relatively flexible bonds between the tetrahedra (see above) leading to lower pressure
rotation of the latters.

After this rotation, almost all Bg modes increase rapidly, which is a logical consequence of the deformations
of $ZnF_4$ tetrahedra on the (001) plane and of the decrease of the interatomic distances on this plane due to
the \textbf{a} decrease and the $ZnF_4$ rotation. It should be noted that the $ZnF_4$ rotation vibration mode
Ag at 138 $cm^{-1}$ and 202 $cm^{-1}$ have smaller slope after $ZnF_4$ rotation, which is comprehensible
because the constraint on the rotational movement should be more important before the relaxation due to the
$ZnF_4$ rotation.

{\bf Singularities at 10 GPa} : Around this pressure, most lines suddenly disappear (see Figure 2) and almost
all observed frequencies are independent of pressure. The samples released from the pressure range 10-16 GPa
can still recover their original phase.

{\bf Singularities at 17 GPa} : All lines disappear except for the Eg mode at 464 $cm^{-1}$ (undergoing a
constant frequency decrease in all the pressure range with a small slope) and the Bg mode at about 120
$cm^{-1}$. These two intense lines under high pressure remain unchanged when the mono-crystal samples are
released from pressure, meaning that this high pressure structure transformation around 17 GPa is irreversible.
This irreversible transformation has been also observed in $YLiF_4$ at much higher pressures (around 30
GPa)\cite{Wang1}.

\section{Conclusion}

In summary, we have presented the experimental results of the pressure effects on the lattice vibration and
Raman spectrum of $CaZnF_4$ up to 17 GPa at which the crystal undergoes an irreversible structural
transformation. Some Raman anomalies in the pressure dependence of Raman frequencies have been observed and
analyzed according to the atomic movements of different vibration modes. No clear evidence of phase transition
has been observed below 17 GPa.

With regard to the mechanism of the possible structure transformations of $CaZnF_4$ under high pressure, no
precision can be given for the time being. All the mechanisms including the rotation and deformation of the
$ZnF_4$ and $CaF_8$ polyhedra could be invoked. It is worth noticing that the internal binding of the $ZnF_4$
tetrahedra is stronger than the external binding in view of the much stronger internal force constant
$A_{Zn-F}\sim 190 N/m$ than the external one $A_{Ca-F}\sim 70 N/m$. So if there is a lattice distortion up to
17 GPa, it can be expected that the polyhedra $CaF_8$ would be more deformed than the $ZnF_4$ tetrahedra which
subsist above 17 GPa because there are still two intense lines, Eg (464 $cm^{-1}$) and Bg (120 $cm^{-1}$),
corresponding to their vibration.

\vspace{2cm}

{\huge Figure captions}

\begin{enumerate}

\item (a) unit cell of the fluoride scheelite $CaZnF_4$ with space group $C_{4h}^6$, showing the spatial
arrangement of individual $ZnF_4$ tetrahedra (b); (c) projection of the $ZnF_4$ tetrahedra on the (001) plane
and evidence of a $CaF_8$ double tetrahedron with the $Ca$ ion in the z=0 plane.

\item Pressure influence on the Raman spectra of $CaZnF_4$. The top spectrum is from a sample released from
above 17 GPa and different from that of the original samples (on the bottom).

\item Pressure dependence of the observed Raman frequencies of $CaZnF_4$. The lines are only guides for the
eye.

\end{enumerate}

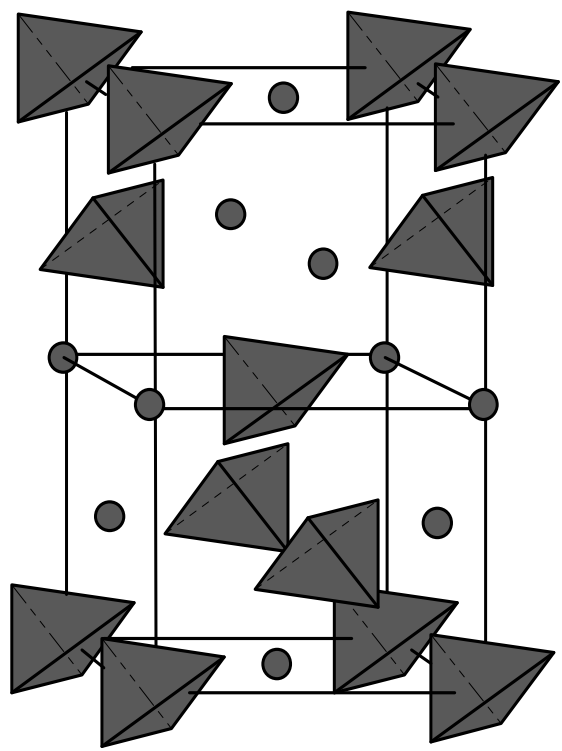
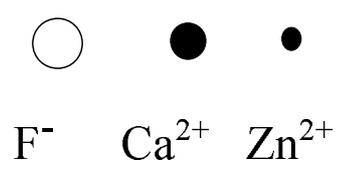
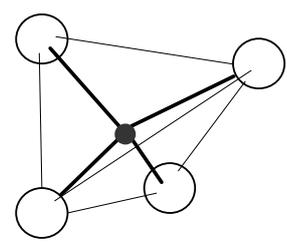
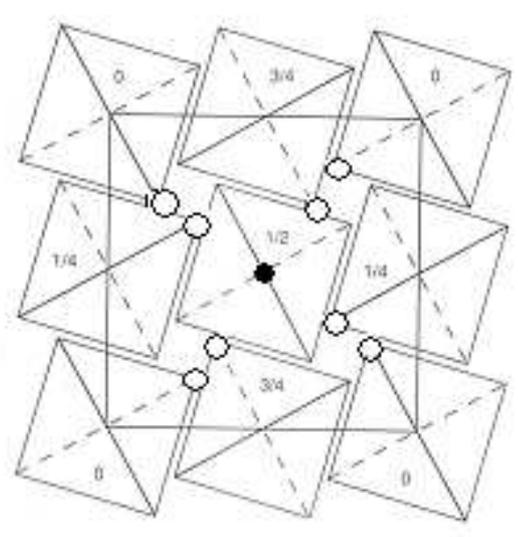
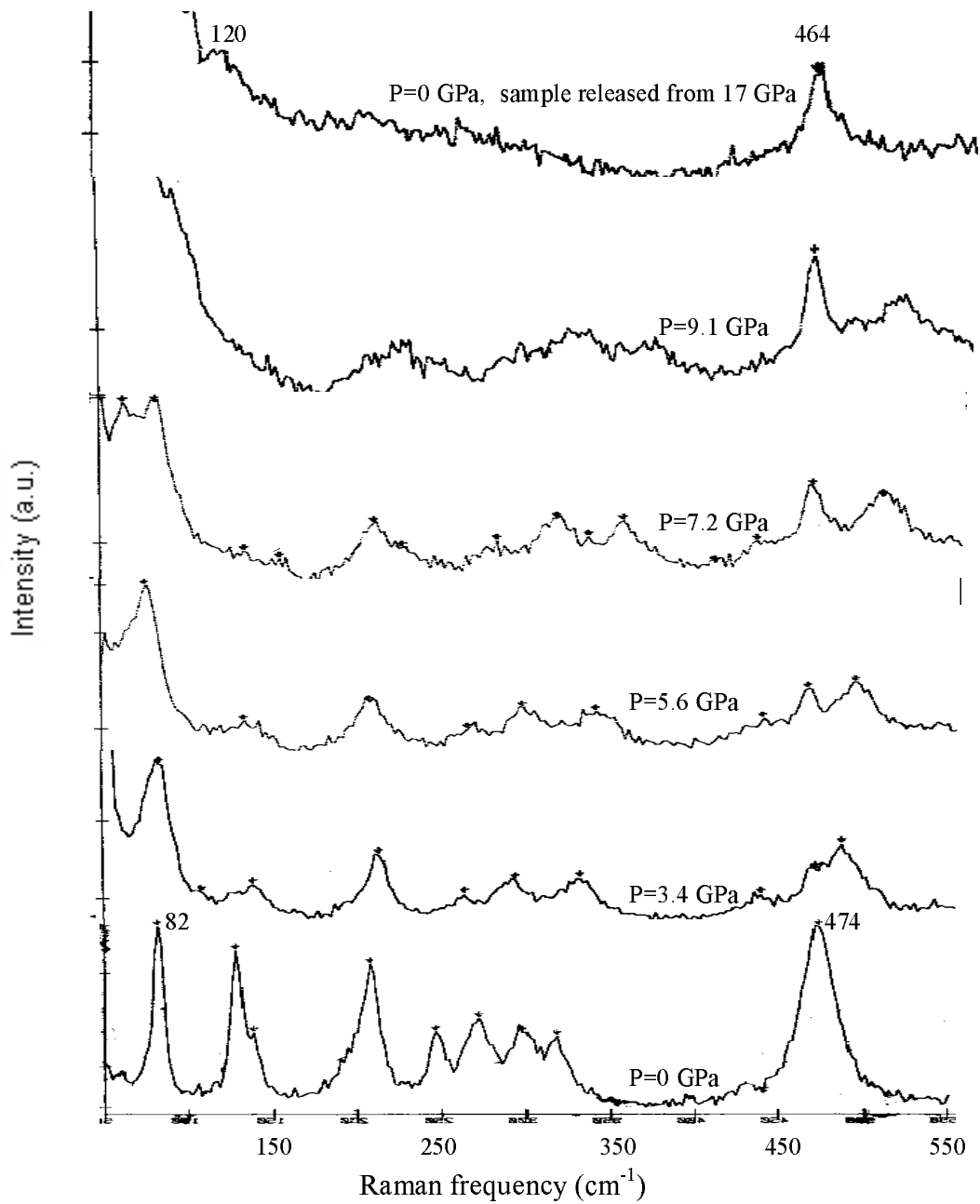
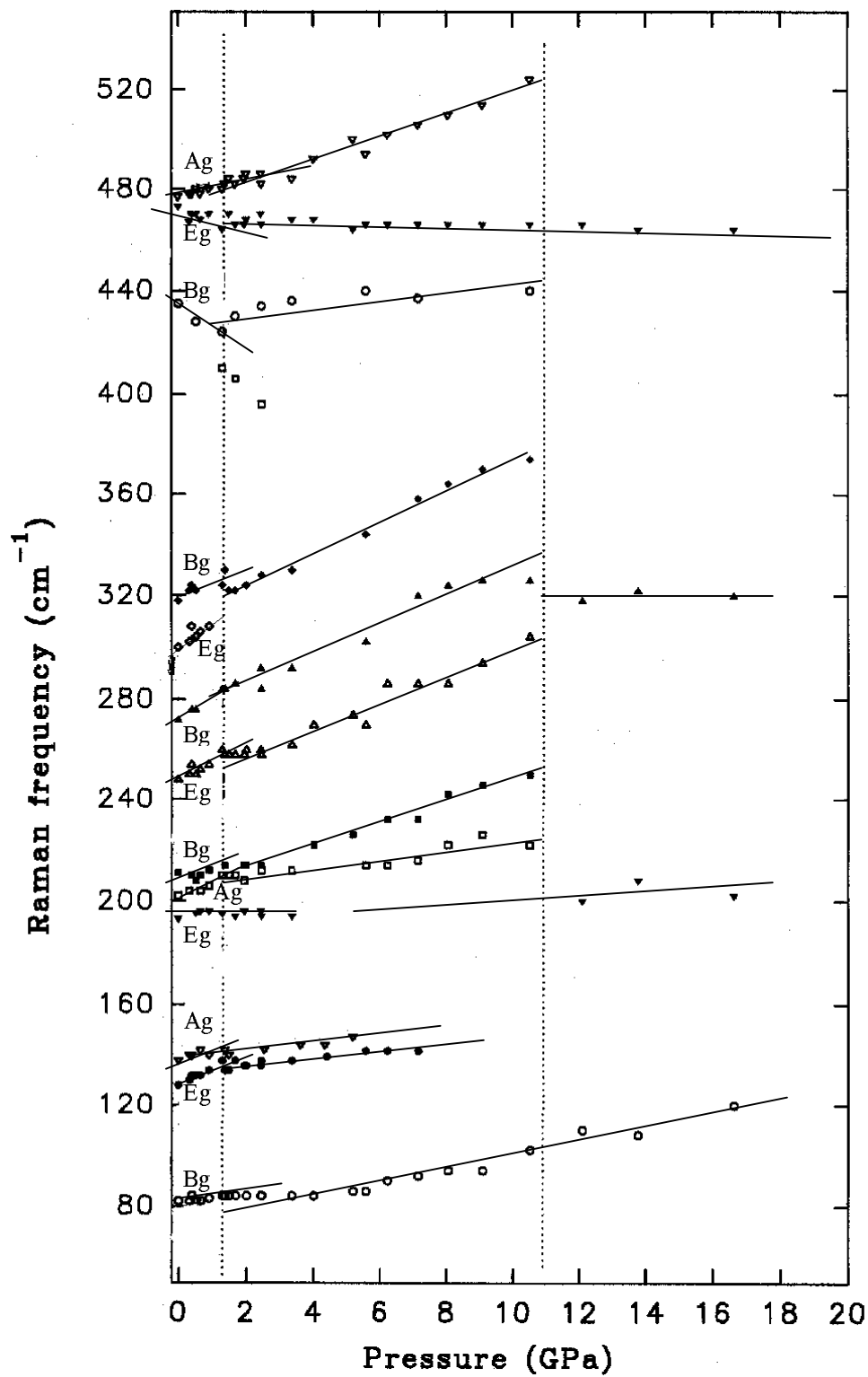
\end{document}